\documentclass[natbib=false,manuscript]{acmart}

\AtBeginDocument{%
  }

\setcopyright{acmlicensed}
\copyrightyear{2024}
\acmYear{2024}
\acmDOI{XXXXXXX.XXXXXXX}

\acmConference[ICTD '24]{the 2024 International Conference on Information \& Communication Technologies and Development}{December 9-11, 2024}{Nairobi, Kenya}
\acmISBN{978-1-4503-XXXX-X/18/06}

\RequirePackage[
  datamodel=acmdatamodel,
  style=acmnumeric,
  ]{biblatex}

\begin{document}

\title{Dukawalla: Voice Interfaces for Small Businesses in Africa}

\author{Elizabeth A. Ankrah}
\email{eankrah@uci.edu}
\orcid{0000-0002-2538-929X}
\affiliation{%
  \institution{University of California, Irvine}
  \country{USA}}

\author{Stephanie Nyairo}
\email{snyairo@microsoft.com}
\orcid{0000-0001-7847-4116}
\affiliation{%
  \institution{Microsoft Research Africa}
  \country{Kenya}}

\author{Mercy Muchai}
\email{mercy.muchai@microsoft.com}
\affiliation{%
  \institution{Microsoft Research Africa}
  \country{Kenya}
}

\author{Kagonya Awori}
\email{Kagonya@google.com}
\orcid{0000-0003-1248-4500}
\affiliation{%
 \institution{Google}
  \country{Kenya}
}

  \author{Millicent Ochieng}
  \email{mochieng@microsoft.com}
  \orcid{0000-0003-4769-7039}
\affiliation{%
 \institution{Microsoft Research Africa}
  \country{Kenya}}

  \author{Mark Kariuki}
  \email{markmuigai@gmail.com}
\affiliation{%
 \institution{University of Nairobi}
  \country{Kenya}}

\author{Jacki O'Neill}
\email{jaoneil@microsoft.com}
\orcid{0000-0002-3065-0579}
\affiliation{%
 \institution{Microsoft Research Africa}
  \country{Kenya}}

\renewcommand{\shortauthors}{Ankrah et al.}

\begin{abstract}
Small and medium-sized businesses (SMBs) often struggle with data-driven decision-making due to a lack of advanced analytics tools, especially in African countries where they make up majority of the workforce. Though many tools exist they are not designed to fit into the ways of working of SMB workers who are mobile-first, have limited time to learn new workflows, and for whom social and business are tightly coupled. To address this, the Dukawalla prototype was created. This intelligent assistant bridges the gap between raw business data and actionable insights by leveraging voice interaction and the power of generative AI. Dukawalla provides an intuitive way for business owners to interact with their data, aiding in informed decision-making. This paper examines Dukawalla's deployment across SMBs in Nairobi, focusing on their experiences using this voice-based assistant to streamline data collection and provide business insights.
\end{abstract}





\maketitle

\section{Introduction}
In an era of data-driven decision-making, small and medium-sized businesses (SMBs) are typically at a disadvantage compared to larger corporations which can leverage sophisticated analytics tools. SMBs typically lack access to technological solutions that align with their unique operational contexts. This disparity is felt keenly in African countries, where SMBs account for approximately 80\% of the continent's workforce \cite{mwarari2013factors} . Here the challenge lies not just in adopting new technologies but in integrating them into the social fabric of these businesses, which plays such a crucial role in their success, a concept known as socio-tecture \cite{awori_its_2022}.

In response to these challenges, we created the Dukawalla prototype, an intelligent Large Language Model (LLM)-based assistant designed to bridge the gap between raw business data and actionable insights for SMB owners. By leveraging the power of generative AI, specifically LLMs, in a conversational format, our solution aims to provide a simple and intuitive way for business owners to interact with their data, fostering reflection on business performance and facilitating informed decision-making. This paper investigates the deployment of Dukawalla across seven SMBs in Nairobi. We explore their experience using this voice-based assistant to streamline data collection and provide insights. 

\section{Related Work}
The importance of cultural context in technology adoption, particularly for SMBs in the Global South, cannot be overstated. Avgerou \cite{avgerou_information_2008} emphasized the need for information systems research to consider local social and cultural conditions, especially in developing countries. Whilst the ubiquity of mobile devices in SMB operations, particularly in developing economies, presents both opportunities and challenges for data management. Nah et al \cite{nah_value_2005} discussed the potential of mobile business applications to provide real-time insights to decision-makers. However, they also noted the limitations of small screens and mobile interactions for complex data visualizations and data input errors.

The emergence of AI and conversational interfaces has introduced new possibilities for supporting reflection in business contexts and offer a promising solution to these challenges. As Følstad and Brandtzæg \cite{folstad_chatbots_2017} note, chatbots can serve as reflection partners, helping users explore and make sense of their data through natural dialogue. This approach aligns with Schön's concept of "reflection-in-action" \cite{Schon_ReflectivePractitioner_1983}, where practitioners think about their actions while performing them. Building on this, recent work by Lee et al. \cite{Lee_WeBuildAI_2019} demonstrates how AI-powered systems can enhance the reflection process by highlighting patterns in financial data, asking probing questions, providing contextual information, and suggesting alternative perspectives. In Dukawalla, by combining voice interaction with an AI-powered system, we aim to create a more intuitive and engaging data management experience for SMB owners. 

\section{Method}
In this paper, we report our findings from the qualitative two week field deployment of Dukawalla. This study was approved by the institutional ethics review board, and consent was obtained from the participating SMBs. Participating SMBs were compensated with a KES 30,000 voucher (KES 10,000 for every week they participated in the study). We recruited seven businesses, and data collection spanned 9 weeks from August 2023 to October 2023. Each SMB sold and bought physical goods or provided a service, had fewer than 300 employees, and used mobile money services (which is the norm in Kenya). The regulatory and institutional framework in Kenya defines small and medium enterprises or small and medium businesses as those employing less than 50 workers \cite{NationalCouncil2012}, whilst Microsoft defines SMBs as businesses with 1-300 employees\cite{Gordon2022}. We advertised the study through digital flyers, face-to-face recruitment in Nairobi marketplaces and snowball sampling. Prior to field deployment, researchers conducted a one-week rapid ethnography with each business to understand its context, culture, and values. The results of this initial fieldwork were used to identify an employee who would be the primary user of the prototype during the field deployment. 

\subsection{System Design and App Flows}
\begin{figure}[h]
    \centering
    \includegraphics[width=1\textwidth]{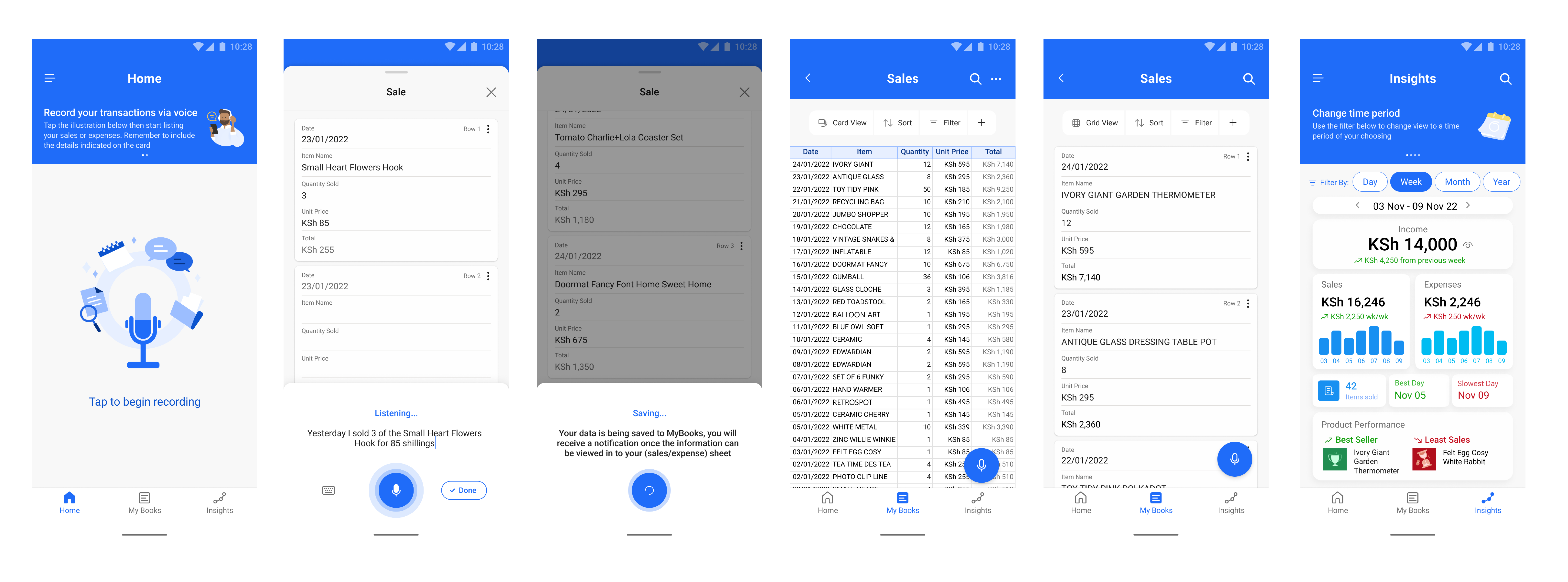} 
    \caption{Mobile app screens for Dukawalla showing the voice recording flow, sales data in My Books, and the infographics in Insights. \href{https://www.microsoft.com/en-us/research/uploads/prod/2024/11/Dukawalla-ICTD-AppScreens.png}{View more screens}}
    \label{fig:Dukawalla_AppScreens}
\end{figure}

Inspired by prior research that found that SMBs lack usable and actionable data management tools for mobile and low-resourced environments, we developed this prototype to explore how a voice and mobile first LLM-based tool could help SMBs in their daily business. Dukawalla supports SMBs in collating their data and gleaning business insights through the incorporation of three main features: 1) \textbf{Voice-enabled assistance}: allowing participants to record data using natural language in an accessible, hands-free method using speech-to-text recognition; 2) \textbf{Language Understanding with Generative AI}: Using LLMs to structure the transcription from the Speech-to-Text to CSV data that would then be stored within their business books;  3) \textbf{Bite sized visualisations: } This feature collated the participants' business data captured via voice or text into a mobile friendly view and generated visualizations of their insights (see Fig. \ref{fig:Dukawalla_AppScreens}) (\href{https://www.microsoft.com/en-us/research/uploads/prod/2024/11/Dukawalla-ICTD-AppScreens.png}{see additional app screens}).
\subsection{Field Deployment}
One employee in each business was given an android mobile phone with the Dukawalla prototype installed, they were asked to use the prototype for two weeks. In the first week, researchers were present to observe interaction with the tool and answer any questions or technical issues. Researchers interviewed participants about their expectations and initial response to the application. In the second week, participants used the tool alone, the week concluded with an exit interview between the researcher and participants. During this session, participants were asked to discuss their experience with the key functionalities and how it impacted their work, data management, and decision-making process.

\section{Findings}
In this section, we report on the results of the  Dukawalla deployment. 

\subsection{Use of Voice: Assessing Language Models and Understanding}
Prior to receiving Dukawalla, participants were hopeful about its impact on their work believing that it would reduce the burden associated with manually inputting data. They hoped that Dukawalla would help to save time when recording data especially in the field. B02\_CEO shared, ``\emph{the voice, even if I haven't used it, I think it can come in very handy because you just talk and in two seconds you\textquotesingle re done. Sometimes you may not have time to key in and type.'' }

Yet despite the excitement in actual use they encountered challenges using Dukawalla to effectively support data collection. For example, participants found it challenging to determine the appropriate times for recording data via voice. As B01\_Admin says  \emph{``You can't start recording, and the clients are waiting to be served. That is a bit of a challenge.'' }. This was echoed by other participants as there was a tension between interacting with the tool and interacting with their customers. As  B01\_Employee shared ``\emph{They (A customer) are like, why are you saying the things that I have bought?''}. Previous research on socio-tecture found that businesses prioritise relational over transactional approaches and using Dukawalla was felt to interfere with this. Others found that in practice they were simply uncomfortable with using voice at work, as B05\_Accountant stated \emph{``I don't know why but I did not try the voice... I think it's weird to me.}'' 

In addition to these social challenges, participants also experienced environmental issues, for example noise and chatter from the surroundings influenced their ability to record. During observation, B02\_CEO was attempting to record some data as their neighbour was speaking French, this caused the recording to fail as the speech detection system could not identify which language to transcribe in. This pointed to a broader challenge as participants frequently codemixed between Swahili, English, and other local dialects, however speech recognition is still largely monolingual.  As a result, the model was not always able to understand or recognize certain words such as ``\emph{Dhania}'', the Swahili word for coriander that B01 sold. This meant that researchers and engineers had to provide updates during the intervention to teach the language model to recognize cultural and organizational idiolect.

In another instance, participants were forced to change their phrasing to align better with the models understanding, for example when recording weight of certain product sold a B01\_Admin stated, \textit{“Yeah, I have sold spinach 0.4 Kg at Ksh...”} However, the system could not determine how to log this measurement data forcing participants had to change their phrasing to 400 grams for the sale to be recorded.

\subsection{Unstructured to Structured data: Assessing Language Understanding}
The LLM performance in the prototype was influenced by two factors the first being how well the speech model transcribed participants speech (see above). The second being the way we designed the interaction to create structured CSV data from unstructured speech. The prototype was customized to have 4 mandatory columns for a sale; date, item, unit sold, and unit price. The LLM structured the speech to text data from the user based on these predetermined columns. However, in practice a there were a number of errors arising from limited understanding of the contextual use of language. For example, the colloquial term ``bob'' is used to denote currency, users would record a sale and say ``sold X item for 120 bob''.  However, the LLM did not recognise this data as currency and would instead return an error or assign it to units sold. Another common example arose from users recording a sale by saying e.g. ``I sold x at one fifty.'' This presented three options for language understanding 1) user sold X at 1:50 (time) 2) User sold X at 150 (KSh, expected currency) 3) user sold X at 1.50 (USD, unexpected currency). For those familiar with spoken language in Kenya the automatic assumption would be (2) as commonly both the currency and the definitive numerical ('hundred') are dropped. Lacking good regional understanding of language use, the LLM often structured the data to the wrong fields. 

\section{Discussion}
In this research we identified three challenges of using voice interfaces to record business data. These include 1) the complexities of using speech in interaction, especially when prioritizing relationships over transactions, 2) the complexities of speech as an input in particular that people naturally codemix and use colloquial terms - either regional or business specific. However currrently ASR models tend to be monolingual and lack data about local contexts, leading to more errors; 3) the complexities of designing an app to support voice interaction for structured data extraction. Whilst huge advances have been made in LLMs as regards natural language prompting in text, the same cannot be said for prompting in voice. Further, users tend to be more familiar with designing their input for text systems. Voice prompts tend to be more free flowing, and combined with the challenges inherent in many ASR systems, we needed to providing some scaffolding for consistent data which we could generate insights from. Even so, there often errors in structuring the data.
Voice presents an opportunity for SMBs to digitize their data and save time with their record keeping. However key challenges that arose from the testing of the Dukawalla prototype need to be addressed for voice interactions to become more seamless and intuitive as limitations with hardware, speech data, and language data resulted in the environmental, social, cultural, contextual and speech model and LLM challenges identified in this paper.

\begin{acks}
We thank the SMB participants in this work for their valuable time and input. We additionally thank the Microsoft team for their continued support. The research reported in this paper was supported by the National Science Foundation Graduate Research Fellowship under Grant No. DGE-1839285. Any opinions, findings, conclusions, or recommendations expressed in this material are those of the author(s) and do not necessarily reflect the views of the National Science Foundation,  .
\end{acks}

\printbibliography


\end{document}